\documentclass[article]{jfm}

\usepackage{graphicx}
\usepackage{newtxtext}
\usepackage{newtxmath}
\usepackage{natbib}
\usepackage{hyperref}
\usepackage{indentfirst}
\usepackage[singlelinecheck = false, width=\linewidth, justification=justified]{caption}
\hypersetup{
    colorlinks = true,
    urlcolor   = blue,
    citecolor  = black,
}

\newcommand{\RomanNumeralCaps}[1]
\linenumbers

\title{Thermal Capillary Wave Growth and Surface Roughening of Nanoscale Liquid Films}

\author{Y. Zhang\aff{1}
  \corresp{\email{Y.Zhang.134@Warwick.ac.uk}},
  J. E. Sprittles\aff{2}\corresp{\email{J.E.Sprittles@Warwick.ac.uk}},
 \and D. A. Lockerby\aff{1}\corresp{\email{D.Lockerby@Warwick.ac.uk}}}

\affiliation{\aff{1}School of Engineering, University of Warwick, Coventry CV4 7AL, United Kingdom
\aff{2}Mathematics Institute, University of Warwick, Coventry CV4 7AL, United Kingdom}

\begin{document}
\maketitle

\begin{abstract}
The well-known thermal capillary wave theory, which describes the capillary spectrum of the free surface of a liquid film, does not reveal the transient dynamics of surface waves, e.g., \,the process through which a smooth surface becomes rough. Here, a Langevin model is proposed that can capture these dynamics, goes beyond the long-wave paradigm which can be inaccurate at the nanoscale, and is validated using molecular dynamics simulations for nanoscale films on both planar and cylindrical substrates. 
We show that a scaling relation exists for surface roughening of a planar film and the scaling exponents belong to a specific universality class.
The capillary spectra of planar films are found to advance towards a static spectrum, with the roughness of the surface $W$ increasing as a power law of time $W\sim t^{1/8}$ before saturation. However, the spectra of an annular film (with outer radius $h_0$) are unbounded for dimensionless wavenumber $qh_0<1$ due to the Rayleigh-Plateau instability. 
\end{abstract}

\section{Introduction}\label{sec1}
Surface roughening due to randomness is ubiquitous in nature, and a problem spanning many disciplines, e.g., in the propagation of wetting fronts in porous media, in the growth of bacterial colonies, and in atomic deposition during the manufacture of computer chips\,\citep{ka1986}. To allow us to predict and control surface roughening, it is essential to understand how surface morphology develops in time, and this is usually described by scaling relations\,\citep{ka1986,ba1995}. 

A liquid film at rest on a substrate also has a rough, fluctuating surface due to thermally excited capillary waves. These capillary waves are used in experiments to measure properties of liquid-solid systems\,\citep{ji2007,po2015,al2012}. This measuring technique has the advantage of being non-invasive, which is important for soft matter and biological fluids that can be sensitive to external forces. Capillary waves also play an important role in modern theories of surface physics\,\citep{ma2013,ev1981}, and are thought to be critical to the instability of thin liquid films\,\citep{vr1968} where thermal capillary waves are enhanced by disjoining pressure, leading to the film rupture. The roughness created by thermal capillary waves is usually on the scale of nanometres, but it has also been observed optically at the microscale in ultra-low surface tension mixtures\,\citep{aa2004}.

Previous work \,\citep{ji2007,po2015,al2012,ma2013,ev1981,aa2004} has been underpinned by Capillary Wave Theory (CWT), which, from the equipartition theorem (i.e. in thermal equilibrium), 
provides the mean amplitude of each surface mode as a function of wavenumber ($q$). For a planar film (without gravity), the r.m.s. spectral density is given by: 
\begin{equation}\label{eq:ss}
S_{\mathrm{s}}(q)\propto\frac{\sqrt{{{k}_{B}T}/{\gamma}}}{q}\,  ,
\end{equation}
where $k_B$ is the Boltzmann constant, $T$ is temperature, and $\gamma$ is the surface tension. Importantly, (\ref{eq:ss}), has no time dependence; it cannot reveal how a smooth surface develops to a rough one, or how sudden changes in material parameters generate evolution towards new spectra -- as such, we refer to it as the {\em static spectrum}, and denote it using the subscript `s'. Understanding dynamics is also essential as it allows prediction of the time required for a smooth film to reach its static spectrum, thus determining when the adoption of classical CWT is valid. 

Alongside open problems concerning the dynamics of capillary waves on a planar interface is their evolution in a cylindrical geometry. An analysis of this geometry is timely, being driven by state-of-the-art applications, e.g., the use of nanofibres to transport annular films of liquids\,\citep{hu2013} and the manufacture of ultra-smooth optical fibres\,\citep{br2017}.

For the aforementioned dynamic problems, it is natural to seek solutions to the equations of fluctuating hydrodynamics (FH)\,\citep{la1959}. In FH, thermal fluctuations, which drive the capillary waves, are modelled by a stochastic stress (white noise) contribution to the Navier-Stokes equations. A long-wave approximation for thin films, and also for jets, has been used to derive stochastic lubrication equations (SLE) from FH:
the Jet SLE by\, \citet{mo2000}, and the Planar-film SLE by\, \citet{gr2006}, \,\citet{da2005} and \,\citet{du2019}. 
Notably, numerical solutions to the Jet SLE\,\citep{mo2000,eg2002,zhao2020} and the Planar-film SLE \,\citep{gr2006,ne2015,di2016,du2019,sh2019} demonstrated that noise can accelerate the rupture process, in agreement with experimental analyses\,\citep{be2003}.

 Linear stability analysis of the SLE has provided time-dependent capillary wave spectra for both jets and planar films\,\citep{me2005,fe2007,zh2019,zhang2019}. 
Importantly, in the recent article\, \citep{zhang2020}, new SLEs have been derived for both planar and annular films (like those consider here), taking into account the slip effects at the solid-liquid interface, which are well-known to be significant for nanoflows\,\citep{la2005,bo2010}. However, despite their success, the long-wave approximation inherent in each of these SLEs creates restrictions on the wavelengths that can be accurately predicted, which requires the development of a more general method.

The motivation of this work is to understand the time-dependent nature of capillary wave spectra, $S(q,t)$, i.e. the surface roughening process (i) for different types of film (e.g. planar or annular), (ii) with different physics (e.g. with or without liquid slip at the substrate), and (iii) without the limitations of the lubrication approach.  The subject is both of fundamental interest and practical value: creating a single theoretical framework under which the time evolution of thermal capillary waves on films can be studied; and allowing prediction of the time required for a smooth film to reach its static spectrum, thus determining when the adoption of classical CWT is valid.

This paper is organised as follows. In \S\,\ref{sec2}, the molecular models of nanoscale liquid films on substrates are presented, which will be used as virtual nanoscale experiments against which to validate new theories. In \S\, \ref{sec3}, our new Langevin model of capillary wave growth is derived. \S\,\ref{sec4} compares the new model with molecular simulation results and previous experiments, and discusses new findings. In \S\ref{sec5}, we summarise the main contributions of this work and outline exciting future directions of research.

\section{Molecular Dynamics Simulations} \label{sec2}
\begin{figure}
\includegraphics[width=\linewidth]{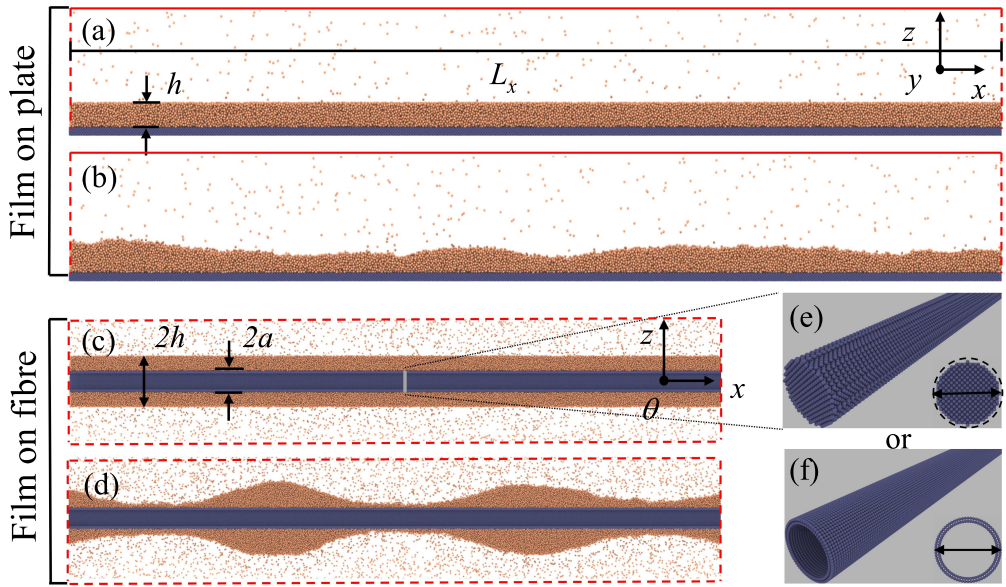}
\caption{Snapshots of a thin liquid film (a section) on a substrate in MD. For planar films, (a) initial  configuration with a smooth surface; (b) surface roughening. For annular films, (c) initial  configuration; (d) beads formed due to the Rayleigh-Plateau instability. Two types of cylindrical substrates are used: (e) Fibre 1, cut out from a bulk of Platinum with fcc structure. (f) Fibre 2, consisting of two concentric surfaces. $L_x$ is the film length, $h$ is the film thickness for a planar film and film radius for an annular film, and $a$ is the fibre radius ($y$ and $\theta$ are into the page).}
\label{fig1}
\end{figure}
We use molecular dynamics (MD) simulations as a benchmark for capillary wave dynamics of nanoscale liquid films (the best proxy for experimental data at this scale), and adopt the popular open-source code LAMMPS\,\citep{pl1995}. All simulation domains contain three phases, with an Argon liquid film bounded by its vapor above and a Platinum substrate below, as shown in figure \ref{fig1} for planar and annular films.

The film is composed of liquid Argon, simulated with the standard Lennard-Jones (LJ) 12-6 potential: $U({{r}_{ij}})=4{{\varepsilon }_{ll}}[(\sigma_{ll}/r_{ij})^{12}-(\sigma_{ll}/r_{ij})^{6}]$, where $ll$ denotes liquid-liquid interactions and $ij$ represents pairwise particles. The energy parameter $\varepsilon_{ll}$ is $1.67\times{10^{-21} }$ J and  the length parameter $\sigma_{ll}$ is $0.34$ nm. The temperature of this system is kept at $T=85$ K or $T^*=0.7 \varepsilon_{ll} /k_B$ (* henceforth denotes LJ units). At this temperature, the number density of liquid Argon is $n_l^*=0.83/\sigma_{ll}^3$. The number density of the vapor phase is $(1/400) n_l^*$. The surface tension of liquid is $\gamma = 1.52\times{10^{-2}} $ N/m  and the dynamic viscosity is $\mu=  2.87\times{10^{-4}}$ kg/(ms)\, \citep{zhang2020}. For a planar substrate, the solid is Platinum made of five layers of atoms, with a face centred cubic (fcc) structure and its $\langle100\rangle$ surface in contact with the liquid. The Platinum number density is $n_s^*=2.60/\sigma_{ll}^3$. For cylindrical substrates, two different atomic structures are considered: the one in figure \ref{fig1}(e) is generated by cutting a cylinder from a large cube of Platinum; the one in figure \ref{fig1}(f) consists of two concentric surfaces, of which the cross section consists of two rings with same number of particles distributed uniformly. All solid substrates are assumed to be rigid, which saves considerable computational cost. The liquid-solid interactions are also modelled by the same 12-6 LJ potential with $\sigma_{ls}=0.8\sigma_{ll}$ for the length parameter. For planar films, three different values of the energy parameter are used, in order to generate varying slip lengths: Case (P1) $\varepsilon_{ls} =0.65\varepsilon_{ll}$, Case (P2) $\varepsilon_{ls} =0.35\varepsilon_{ll}$ and Case (P3) $\varepsilon_{ls} =0.20\varepsilon_{ll}$. For annular films and Fibre 1, Case (A1) $\varepsilon_{ls} =0.7\varepsilon_{ll}$; for Fibre 2, Case (A2) $\varepsilon_{ls} =0.6\varepsilon_{ll}$.
 
The initial dimensions of a planar liquid film (see figure \ref{fig1}(a)) are $L_x=313.90$ nm, $ L_y=3.14 $ nm and $h_0=3.14$ nm; the MD simulations are quasi-2D ($ L_x\gg L_y$) allowing comparison with 2D theory. The initial size of the annular film (see figure \ref{fig1}(c)) has film length $L_x=229.70$ nm and outer radius $h_0=5.74$ nm.  The radius of Fibre 1 is defined by the radius of cylinder, $a_1=2.35$ nm, used to cut the fibre out of a bulk cube of Platinum. Fibre 2 has an outer radius $a_2=2.17$ nm, with spacing 0.22 nm from the inner ring. Solid particles are distributed uniformly with $5^\circ$ spacing.

For the planar case, we separately equilibrate a liquid film with thickness $h_0=3.14$ nm and a vapour are in periodic boxes at $T=85$ K. Then the film is deposited above the substrate and the vapour is placed on top of the film. Because there exists a gap (a depletion of liquid particles) between the solid and liquid, arising from the repulsive force in the LJ potential, it is necessary to deposit the liquid above the substrate by some distance. The thickness of the gap is found to be about $0.2$ nm after the liquid-solid system reached equilibrium so that we choose a deposit distance $d=0.2$ nm. This makes the position of film surface at $z=h_0+d=3.34$ nm initially if the substrate surface has position at $z=0$.

For an annular film, cuboid boxes of liquid and vapour are equilibrated separately in periodic boxes at $T=85$ K. Then an annular film is cut out from the cuboid box with the outer radius at $5.74$ nm and inner radius above the fibre radius with an interval $0.2$ nm. Then the fibre is put into the annular film and vapour is placed to surround the film. Notably, in this case, the position of film surface is still at $h_0=5.74$ nm initially.
 
Periodic boundary conditions (PBC) are applied in the $x$ and $y$ directions of a planar system whilst vapour particles are reflected specularly in the $z$ direction at the top boundary of the planar system. For annular films, PBC are applied in all three directions. After initialization of the simulated systems, the positions and velocities of the liquid and vapour atoms are updated with a Nos\'{e}-Hoover thermostat (keeping the temperature at $T=85$K) and the above boundary conditions.

According to classical theory, the surface of the initially smooth planar film, figure \ref{fig1}(a), should remain smooth indefinitely. However, thermal fluctuations generate surface roughness over a period of time, see figure \ref{fig1}(b), and it is the evolution of this roughness that we study here. The situation for the annular film is more complex, since, as seen in figure \ref{fig1}(d), it can be prone to a Rayleigh-Plateau instability, due to the `pinching' surface tension force generated by the circumferential curvature.

In this work, $h(x,t)$ is the film height, $\delta{h}=h-h_0$ is the surface perturbation from its initial height $h_0$, $\widehat{\delta{h}}$ the perturbation in Fourier space, and $S(q,t)=\sqrt{\langle|\widehat{\delta{h}}|^2\rangle}$ the surface spectrum, where $\langle\, \cdots\rangle$ denotes an ensemble average, and $|\cdots|$ the norm of the transformed variable.
In MD, the liquid-vapour interface ($h$) is defined by the equimolar surface. A discrete Fourier transform of $\delta h$ is performed and surface spectra (presented in \S\,\ref{sec4}) are obtained from the average of a number of independent simulations (65 for planar films and 10 for annular films). 
\section{Langevin Model for Thermal Capillary Wave Growth}\label{sec3}
For nanoflows, where inertia is usually negligible, Stokes flow governs the flow dynamics.  In this regime, for the deterministic setting, linear analyses of free-surface flows are described by equations for the surface perturbation (in Fourier space) of the form:
\begin{equation} \label{eq:stokes}
 \frac{\partial}{\partial t}\widehat{\delta h}+\Omega \,\widehat{\delta h}=0\, ,
\end{equation}
where $\Omega(q)$ is the dispersion relation (the decay rate of a particular mode). The dispersion relation for films depends on the domain geometry, the physics at play, and any approximations adopted. For example, for a planar film with slip, no disjoining pressure, and a long-wave approximation, the dispersion relation is \,\citep{zhang2020}:
 \begin{equation}\label{eqdrpf}
 \Omega(q)=\frac{(h_0^3+3\ell h_0^2)\gamma {{q}^{4}}}{3\mu }\, .
 \end{equation} 
Here $\ell$ is the slip length at the liquid-solid interface and $\mu$ is the viscosity of the liquid. A similar expression is obtained in\, \citep{zhang2020} for the annular film, again, adopting a long-wave approximation.

For nanoscale liquid films, where the Reynolds number is small, Stokes flow is accurate but the long-wave approximation is less valid, particularly as noise can excite short-wavelength perturbations. For planar films with slip, the Stokes-flow dispersion relation was obtained in\, \citep{he2007}, whilst for annular films with slip, we have derived an expression for the first time, with details of the relatively standard derivation in Appendix A.

The main idea in this work is to establish a framework for taking thermal fluctuations into account in modelling films in the {\em general} case (i.e. for whichever film geometry, physics, or modelling approximation we adopt). Knowing the restoring pressure due to surface tension is $\beta\widehat{\delta h}$ ($\beta=\gamma q^2$ for planar films and $\beta=\gamma (q^2-1/h_0^2)$ for annular films), we can rewrite (\ref{eq:stokes}) and add a fluctuating pressure term (white noise) at the same time. This results in a Langevin equation of the form:
\begin{equation}\label{eqn:lan}
\frac{\beta}{\Omega }\frac{\partial \widehat{\delta h}}{\partial t}=-\beta\widehat{\delta h}+\zeta \widehat{N},
\end{equation}
where $\widehat{N}(q,t)$ is a complex Gaussian random variable with zero mean and correlation $\langle | \widehat{N}\widehat{{N'}}| \rangle=\delta( q-q')\delta(t-t')$, and $\zeta$ is the noise amplitude. Since (\,\ref{eqn:lan}) is an Ornstein–Uhlenbeck process, $\zeta$ is determined straightforwardly by considering the surface at thermal equilibrium, where $\langle|\widehat{\delta h}|^2\rangle_\mathrm{s}=S_\mathrm{s}^2=\frac{\zeta^2\Omega}{2\beta^2}$. Thus, we must have
\begin{equation}\label{eqn:noise}
\zeta=\sqrt{\frac{2}{\Omega}}\beta S_\mathrm{s},
\end{equation}
where for planar and annular films, the CWT gives: 
\refstepcounter{equation}\label{eq:ssab}
\[
 S_{\mathrm{s}}= \sqrt{\frac{{{L}_{x}}}{{{L}_{y}}}\frac{{{k}_{B}}T}{\gamma {{q}^{2}}}}, \,\, \mathrm{ } \,\,  S_{\mathrm{s}}=\sqrt{\frac{{{L}_{x}}}{2\pi h_0}\frac{{{k}_{B}}T}{\gamma ({{q}^{2}}-1/{h_0^2})}}  \tag{\theequation a,b},
\]
respectively\,\citep{zhang2020}. Here $L_y$ is the planar-film length in the $y$ direction with $L_x\gg L_y$, making the simulation of planar films quasi-two dimensional. 

The required time-dependent capillary wave spectra can be obtained directly from (\,\ref{eqn:lan}) (see Appendix B). As we are interested in how thermal fluctuations roughen a surface (i.e., the evolution of a non-equilibrium surface to its thermal equilibrium), the initial condition of the surface is assumed to be smooth, though our theory is general to consider any kinds of initial conditions (see Appendix B). As you will see, a smooth interface allows us to extract the maximum time for a non-equilibrium liquid surface to reach its thermal equilibrium, which provides a useful guideline either for computational or experimental investigation of non-equilibrium surfaces. For an initially smooth surface ($S(q,0)=0$), the spectra are:
\begin{align}\label{eqls}
S(q,t)=S_\mathrm{s}\sqrt{\left( 1-{{e}^{-2{{\Omega }}t}} \right)}.
\end{align}

The power of this Langevin model is its generality: to find the time-dependent spectrum for the linear treatment of any Stokes-flow film, all that is required is to substitute the appropriate static spectra and dispersion relation into \eqref{eqls}. For example, substituting \eqref{eqdrpf} and (\ref{eq:ssab}a) into \eqref{eqls} generates exactly the spectra derived in \,\citep{zhang2020} for a planar film, with slip, without disjoining pressure, and using a long-wave approximation (i.e. the SLE). It also offers the opportunity of improving on such SLE predictions by adopting more accurate dispersion relations, such as those utilising Stokes flow (see Appendix A), or adding additional physics without having to always return to the full equations of FH, and performing an asymptotic analysis. 

This Langevin model also naturally bridges the gap between the growth of capillary waves to the static spectrum which is our focus here, and the relaxation of capillary wave correlations after the free surface reaches the static spectrum, widely studied in previous work\,\citep{ji2007,po2015,al2012,aa2004}. With \eqref{eqn:lan} and using the It\^{o} integral\,\citep{di2016, me2005}, the correlation of interfacial Fourier modes is found to be for planar films 
\begin{equation}\label{eqnco}
   \left\langle  \widehat{\delta h}(q,t){{\widehat{\delta h}}^{*}}(q,{t}')  \right\rangle =\left\langle {{\left| \widehat{\delta h}(q,0) \right|}^{2}} \right\rangle {{e}^{-\Omega(t+{t}')}} -\frac{{{L}_{x}}}{{{L}_{y}}}\frac{{{k}_{B}}T}{\gamma {{q}^{2}}}\left[ {{e}^{-\Omega(t+{t}')}}-{{e}^{-\Omega \left| t-{t}' \right|}} \right] 
\end{equation}
Here, the asterisk denotes a conjugate value and, $\langle {{| \widehat{\delta h}({q},0) |}^{2}} \rangle=S(q,0)^2$ is the initial spectra of the surface. Assuming the initial surface is smooth, and with $t=t'$, (\ref{eqnco}) is simplified to (\ref{eqls}). On the other hand, assuming the initial condition is at the state of the static spectrum, (\ref{eqnco}) is reduced to
\begin{equation}\label{eqnre}
\left\langle  \widehat{\delta h}({{q}},t){{\widehat{\delta h}}^{*}}({{q}},t')  \right\rangle =\frac{L_x}{L_y}\frac{{{k}_{B}}T}{\gamma {{q}^{2}}}{{e}^{-\Omega\left| t-t' \right|}},
\end{equation}
which is the relaxation dynamics of capillary waves studied previously\,\citep{ji2007,po2015,al2012,aa2004}. 
\section{Results and Discussions}\label{sec4}
\subsection{Spectra of Planar Films}
\begin{figure}
\centering
\includegraphics[width=\linewidth]{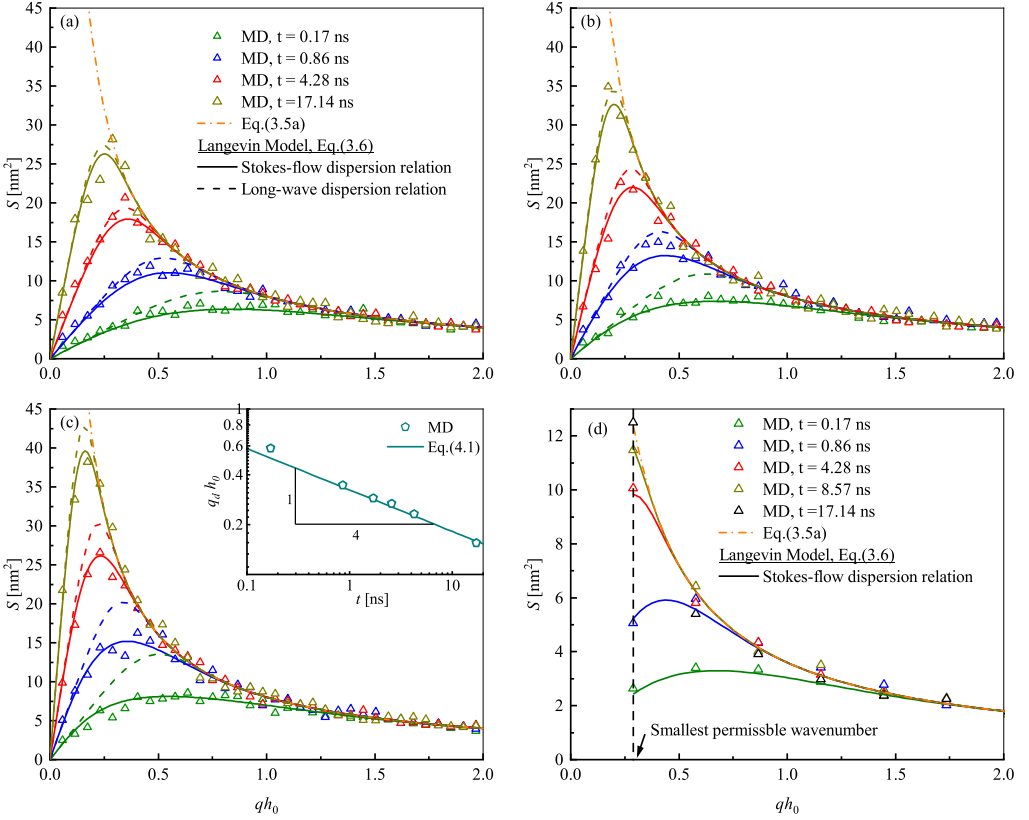}
\captionsetup{justification=justified, singlelinecheck=off} 
\caption{(a-c) Evolution of capillary spectra of a (long) planar film for increasing slip length. A comparison of spectra extracted from MD results (triangles), and Langevin model with Stokes-flow dispersion relation (solid lines) or with long-wave dispersion relation (dash lines) at four different times, along with the static spectrum (dash-dot line). (a) $\ell=0.68$ nm, (b) $\ell=3.16$ nm, (c) $\ell=8.77$ nm. The effective thickness of the film $h_0=2.90$ nm and film length is 313.9 nm. Inset of (c) shows how the dominant wavenumber $q_d$ decreases with time. (d) evolution of capillary spectra for a short film (62.8 nm) with other settings the same with (b).}
\label{fig2}
\end{figure}
We now compare the proposed Langevin model directly to MD data. Figures \ref{fig2}(a-c) show spectra of (long) planar films with three different slip lengths. The first thing we note is that the spectra are, indeed, time dependent, and only gradually approach the static spectrum. One can see that the transient characteristics of the spectra are strongly influenced by the slip length, which is controlled in the MD indirectly by the solid-liquid interaction potential (Appendix C provides details on how this parameter, and the effective film thickness, are extracted from independent MD simulations for use in the Langevin model, see the caption of figure \ref{fig2} for values).

From figures \ref{fig2}(a-c), the MD spectra compare remarkably well with the Langevin model when a Stokes-flow approximation to the dispersion relation is adopted (solid lines) for all slip lengths and at all times. In contrast, the Langevin model with a dispersion relation derived from a long-wave approximation (dashed lines) -- equivalent to the SLE of \,\citet{zhang2020} --  is only accurate (i) when slip lengths are small relative to the film thickness (i.e.\ not for the case in figure \ref{fig2}(c)) and (ii) only in the later stages of capillary wave growth where the dominant (dimensionless) wavenumber $q_dh_0$ (the one with peak amplitude) becomes much smaller than unity (i.e.\ when the wavelength becomes large), as discovered and detailed in\,\citet{zhang2020}. Thus,  the new Langevin model developed here allows us to go beyond the long-wave paradigm.

The dominant wavenumber is seen to decrease with time and $q_d$ can be estimated from the dynamic spectrum \eqref{eqls}, by finding the spectrum's maximum $\partial {S}/\partial q|_{q=q_d}=0$. Adopting the long-wave approximation for the dispersion relation \eqref{eqdrpf}\, allows analytical results to be obtained \citep{zhang2020}:
 \begin{equation}\label{eqqd}
 q_d\cong\left[\frac{15}{8}\frac{\mu}{\gamma (3\ell h_0^2+h_0^3)}\right]^{\frac{1}{4}}t^{-\frac{1}{4}} .
\end{equation}
As can be seen from the inset of figure \ref{fig2}(c), this prediction agrees well with the MD.
\subsection{Roughness of Planar Films and Their Universality Class}
For the free surface considered here, the roughness of the film, $W$, can be defined in terms of the evolving surface spectrum from Parseval's theorem:
\begin{figure}
\centering
\includegraphics[width=\linewidth]{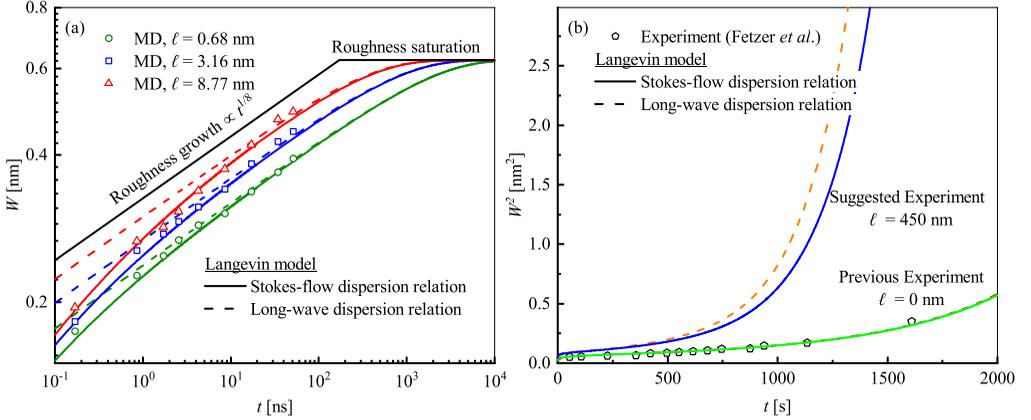}
\caption{Slip effects on surface roughening of a planar film. (a) a comparison made among MD results (symbols), Langevin model with Stokes-flow dispersion relation (solid lines) or with long-wave dispersion relation (dash lines). (b) a comparison of Langevin model with previous experiments\,\citep{fe2007} of a rupturing film without slip but with effects of disjoining pressure. A further experiment with large slip is suggested.
\label{fig3}}
\end{figure}

\begin{equation}
W(t)=\sqrt{\frac{1}{{{L}_{x}}}\left\langle\int_{0}^{{{L}_{x}}}{{{\left( \delta h \right)}^{2}}\,dx}\right\rangle}=\sqrt{\frac{1}{2\pi {{L}_{x}}}\int_{\frac{2\pi}{{{L}_{x}}}}^{\frac{2\pi N}{{{L}_{x}}}}{S^2\, dq}}\, ,
\end{equation}
where $N$ is the number of bins used to extract the surface profile from MD simulations, which provides an upper bound on the wavenumbers that can be extracted. A quick inspection of the MD results presented in figure \ref{fig3}(a) (symbols) reveals that, approximately, the roughness grows with some power law in time, which motivates the use of scaling relations to study surface roughening, as considered previously for the interface roughening between two immiscible inviscid gases \citep{fl1995}. In other words, this opens up the remarkable possibility of obtaining a simple parametrisation for this complex roughening process that aligns the process to seemingly unrelated physical phenomena.

Scaling relations for surface roughness can be summarized by\,\citet{ba1995} 
\begin{equation}
 W\sim L^{\alpha} f(t/L^m),
\end{equation}
where $L$ is the system size, $f(v)=v^{\kappa}$ for $v\ll 1$ (during roughness growth), and $f(v)=1$ for $v\gg 1$ (at roughness saturation; which is not reached in the MD results of figure \ref{fig3}). The time to transition, between roughness growth and saturation, scales with $t_s \sim L^m$. The three exponents ($\alpha$, $m$ and $\kappa$) define a universality class, and are here related by $\kappa = \alpha /m$. 

For the planar film, $\alpha$ can be obtained by considering the surface at saturation, i.e. from the static spectrum given in (\ref{eq:ssab}a), assuming $L_y$ is fixed: $W\sim {L_x}^{{1}/{2}}$. An upper estimate on the transition time, between growth and saturation, can be estimated from the inverse of the dispersion relation at the largest permissible wave length ($q={2\pi}/{L_x}$). For this it is reasonable to use the long-wave approximation, \eqref{eqdrpf}, to find $t_s \sim {L_x}^{4}$, and thus $W\sim t^{1/8}$. In summary, we find the exponents $\alpha =1/2$, $m=4$ and $\kappa=1/8$, assuming, as we have done, long-wave dominated roughness.
 
 The MD results in figure \ref{fig3}(a) indicate that, indeed, $W\sim t^{{1}/{8}}$; this scaling is more apparent at later times, but before saturation, when the roughness is characterised by long wavelengths. This precise scaling, as well as the anticipated roughness saturation, is confirmed by the Langevin model with a long-wave approximation to the dispersion relation (the dashed lines). A closer agreement with MD at earlier times, when the roughness has a shorter characteristic wavelength, is provided by a Stokes-flow dispersion relation (solid lines), but this model does not permit the simple extraction of power laws. The results also show that enhanced slip accelerates the roughening of the surface, but does not alter the final saturated value. 

Interestingly, the new analysis enables us to see that the exponents we find for the surface roughening of a planar film using the long-wave dispersion relation (i.e. the planar-film SLE by\,\citet{gr2006,zhang2020}) are the same with those for surface roughening of atomic depositions in molecular beam epitaxy (MBE)\,\citep{ba1995,krug1997origins}. Thus the two distinct physical problems belong to the same universality class (1/2, 4, 1/8).   

The strong dependency of the transition time on domain length ($t_s\sim L_x^4$) which we have uncovered, explains why in our simulations for a film length $L_x=313.9$ nm this time is of the order of microseconds (see figure \ref{fig3}(a)) and is thus impossible to resolve in MD. For example, for case P2, the transition time $t_s=1/|\Omega|=3389.3$ ns using the long-wave dispersion relation and $t_s=3458.5$ ns using the Stokes-flow dispersion relation, evaluated at the smallest permissible wavenumber, $q=2\pi/L_x$. However for a shorter film with film length 
62.78 nm (other parameters are the same with P2), the transition time is $t_s=5.4$ ns with the long-wave dispersion relation and $t_s=8.2$ ns with the Stokes-flow dispersion relation (with better accuracy). Thus, the complete evolution of capillary waves to the static capillary wave can be realized in MD simulations, which is shown in figure \ref{fig2}(d), but our results have highlighted that care should be taken when interpreting results for larger film lengths where reaching thermal equilibrium (the static spectrum) for the surface is often computationally intractable.
\subsection{Spectra of Annular Films}
\begin{figure}
\centering
\includegraphics[width=\linewidth]{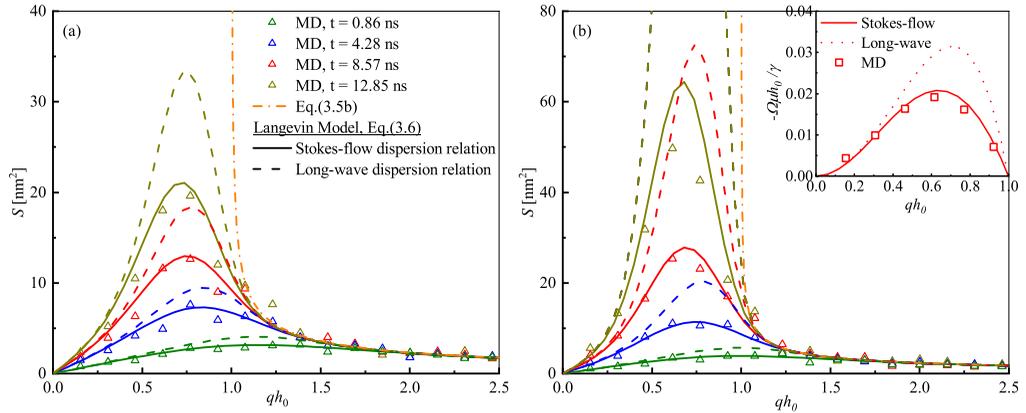}
\caption{Evolution of capillary spectra for annular films; a comparison between MD results (triangles), the static spectrum (dashed and dotted line) and the Langevin model. Dispersion relations used in the Langevin model assuming Stokes flow (solid lines) and a long-wave approximation (dashed lines). Slip lengths (a) Fibre 1, $\ell=0$; and (b) Fibre 2, $\ell=1.18$ nm. The hydrodynamic boundary is at radius $a=2.60$ nm and the initial surface at $h_0=5.74$ nm (see Appendix C for measurement details).
} 
\label{fig4}
\end{figure}

Figure \ref{fig4}  shows the evolving spectra of the capillary waves of annular films.  For wavenumber $qh_0>1$, the MD spectra (triangles) of different times collapse onto the static spectrum, (\ref{eq:ssab}b). However, for $qh_0<1$, the Laplace pressure from the circumferential curvature results in a negative dispersion relation such that the amplitude grows unboundedly until the film ruptures and beads are formed (seen in figure \ref{fig1}(d)).

The surprise finding discussed earlier is that the noise amplitude in the Langevin model appears independent of whether CWT (which assumes disturbances are saturated) or a long-wave approximation (which does not) is adopted. It is therefore interesting to see that the Langevin model compares closely to the MD simulation for the annular film, particularly in unstable regions of the spectra. Note, while the noise amplitude seems independent of the long-wave approximation, the dispersion relation is not, see the inset of figure \ref{fig4}(b); hence the improved agreement when adopting the Stokes-flow dispersion relation, particularly in figure \ref{fig4}(b), which is rather dramatic in the annular case.

\subsection{Connections with Experiments}
Using the parameters found in experiments of polymer systems considered in \citep{ji2007,po2015,al2012} where a static spectrum has to be presupposed to measure the temporal correlations of capillary waves, i.e., (\ref{eqnre}), we calculate the transition time to be hours long -- it is therefore not immediately clear that the assumption of saturation is justified, and this should be confirmed before analysing experimental data.

\citet{fe2007} presented experiments of dewetting polymer films and compared the experimental data with the \emph{no-slip} planar-film SLE\,\citep{gr2006,me2005} to investigate the effects of thermal fluctuations on thin-film dewetting. The high viscosity of experimental liquids makes the time scales for instability growth so slow that AFM can be used to provide spatio-temporal observations. One of the variables they analysed is the roughness of the film surface in experiments, with which we can compare our developed Langevin model.
In their experiments, the dewetting is influenced by disjoining pressure so that the capillary spectrum from Langevin model is slightly modified to consider disjoining pressure $\phi$:
\begin{equation} \label{eqn36}
S({q},t)=\sqrt{{{S}^{2}}({q},0){{e}^{-2\Omega t}}+{{L}^{2}}\frac{{{k}_{B}}T}{\gamma {{q}^{2}}+d \phi /d h\left| _{{{h}_{0}}} \right.}\left( 1-{{e}^{-2\Omega t}} \right)}.
\end{equation}
Here $\phi=\frac{A}{6\pi h_0^3}$ and $A$ is the Hamaker constant. The long-wave dispersion relation considering disjoining pressure is
\begin{equation}
\Omega_\mathrm{LW}=\frac{1}{\mu}\left(\frac{1}{3}h_0^3+\ell h_0^2\right)\left( \gamma {{q}^{4}}+\frac{d \phi }{d h}\left| _{{{h}_{0}}}{{q}^{2}} \right. \right),
\end{equation} 
while the Stokes-flow dispersion relation considering disjoining pressure is
\begin{equation}
 {{\Omega}}_\mathrm{Stokes}=\frac{\gamma {{q}^{2}}+{{\left( d\phi /dh \right)}{\left|_{{h}_{0}} \right.}}}{4\mu q}  \frac{\sinh (2qh_0)-2qh_0+4q\ell{{\sinh }^{2}}(qh_0)}{{{\cosh }^{2}}(qh_0)+{{q}^{2}}{{h_0^2}}+q\ell\left[ 2qh_0+\sinh (2qh_0) \right]}.
\end{equation}

The surface roughness $W$ is thus determined by the spectrum with
\begin{equation}\label{eqn39}
{{W}}=\sqrt{\frac{1}{{{L}^{2}}}\int_{0}^{L}\int_{0}^{L}{{{\left( \delta h \right)}^{2}}}\,dxdy}=\sqrt{\frac{1}{2\pi {{L}^{2}}}\int_{{{q}_{\min }}}^{{{q}_{\max }}}{Sq\,{d}q}}.
\end{equation}
Here one has to think of the spectrum as radially symmetric in the wavenumber space for a two-dimensional surface.

We use the data of roughness from Experiment 1 (Exp. 1) presented in the figure 2 of  \citet{fe2007}. To evaluate (\ref{eqn39}), the values of parameters (film thickness, surface tension, Hamaker constant, viscosity, $q_{min}$, $q_{max}$ and initial condition $S({q},0)$) have to be known. Some of them ($q_{min}$, $q_{max}$, $S({q},0)$) are unavailable from \citet{fe2007}. For others,  \citet{fe2007} provided referenced values but did not provide the fitting values of parameters used to have good fit with experimental data. Therefore, we have to adjust some values of parameters to have best match with the fitting curve in the figure 2 of  \citep{fe2007} to infer what values may have been used by \citet{fe2007}. In summary, the values we use are $h_0= 3.9$ nm, $\gamma=0.045$ N/m, $\mu$ = $2\times 10^4$ kg/(ms), $A = 2\times 10^{-20}$ J, $S({q},0)=0$, 
$q_{min} = 0.42$ nm\textsuperscript{-1} and $q_{max}$ = $0.1q'_d$, where $q'_{d}$ is the constant dominant wavenumber and $q'_{d}=\sqrt{\frac{1}{\gamma}\frac{A}{2\pi h_0^4}}$.

Since there is no slip ($\ell=0$ nm) and the $q'_dh_0$ = 0.0482, which is much smaller than 1, the Langevin model with a long-wave dispersion relation (i.e., Planar-film SLE) can be as accurate as Langevin model with a Stokes-flow dispersion relation. However, polymer films usually have a large slip (up to 1$\upmu$m)\,\citep{fe2005,ba2014} on certain substrates. We thus suggest a experiment using the same polymer film mentioned above, but the film has thickness $h_0$ = 9 nm and a large slip length $\ell = 450$ nm on a substrate. We predict that this would greatly accelerate the roughening and thus dewetting as shown in figure \ref{fig3}(b), which highlights the better accuracy of using the Langevin model with Stokes-flow dispersion relation. 

\section{Conclusion}\label{sec5}
We have investigated the dynamic capillary waves of both planar and annular liquid films at the nanoscale. A Langevin model with a Stokes-flow dispersion relation is able to accurately predict the growth of capillary waves with slip effects, as validated by MD simulations. Though our MD simulations of the evolution of an initially smooth surface is ideal, it may represent the scenario of the melting of nanoscale metal surfaces by laser pulses\,\citep{gon2013}. Our work also provides grounds for carefully evaluating future experiments of thin films that currently rely on Capillary Wave Theory. The quantitative analysis of spontaneous roughening, which is connected to the theory of Universality Classes, allows better understanding of the instability of liquid-vapour or liquid-liquid interfaces\,\citep{vr1968}. Though gravity is not considered in current work as it is usually neglected at the nanoscale, the introduction of gravity to our Langevin model is straightforward for potential applications in larger scales. A topic of future interest will be to investigate how capillary length influences the roughening process. The established relation between capillary spectra and slip also provides a method to measure large slip length such as water films on graphene where a shear-driven method shows considerable statistical errors\,\citep{ku2012}.

\section*{Acknowledgement}
We acknowledge the financial support by the EPSRC (grants EP/N016602/1, EP/P020887/1, EP/S029966/1 \& EP/P031684/1) and the use of Warwick  Scientific Computing Research Technology Platform.
\section*{Declaration of interests}
The authors report no conflict of interest.
\renewcommand{\theequation}{A\arabic{equation}}
\setcounter{equation}{0}
\section*{Appendix A: Stokes-flow dispersion relation for an annular film with slip}
For a liquid film flowing on a fibre, axisymmetric Stokes flow is assumed in the annular film. We use the method in \citep{cr2006} to calculate the dispersion relation analytically, but now assuming we have slip at the liquid-solid interface. The momentum equations are 
\begin{align} \label{dis-mom12}
  & \frac{\partial p}{\partial r}=\mu \left[ \frac{\partial }{\partial r}\left( \frac{1}{r}\frac{\partial (ru)}{\partial r} \right)+\frac{{{\partial }^{2}}u}{\partial {{x}^{2}}} \right], \\
 & \frac{\partial p}{\partial z}=\mu \left[ \frac{1}{r}\frac{\partial }{\partial r}\left( r\frac{\partial w}{\partial r} \right)+\frac{{{\partial }^{2}}w}{\partial {{z}^{2}}} \right]. 
\end{align}
Here $u$, $w$ and $p$ are the radial velocity, axial velocity and pressure, respectively. The mass conservation with incompressible assumption is 
\begin{equation} \label{dis-con3}
\frac{1}{r}\frac{\partial (ru)}{\partial r}+\frac{\partial w}{\partial z}=0.
\end{equation} 
  
In terms of the boundary conditions, we have the slip boundary condition and no-penetration condition at the fibre surface $r=a$ such that 
 \begin{align}
  & w=\ell\frac{\partial w}{\partial r}, \label{dis-bou4}\\ 
 & u=0.\label{dis-bou5}
\end{align}
At the free surface $r=h$, the no-shear boundary condition, for small surface perturbations, is 
\begin{equation}\label{dis-nos6}
\frac{\partial w}{\partial r}+\frac{\partial u}{\partial z}=0,
\end{equation}
and the normal force balance requires (for small surface perturbations)
\begin{equation}\label{dis-nor7}
 -p+2\mu \frac{\partial u}{\partial r}=\gamma \left\{ \frac{{{\partial }^{2}}h/\partial {{z}^{2}}}{{{\left[ {{(\partial h/\partial z)}^{2}}+1 \right]}^{3/2}}}-\frac{1}{h{{\left[ {{(\partial h/\partial z)}^{2}}+1 \right]}^{1/2}}} \right\}.
\end{equation}
Meanwhile, the kinematic condition is
\begin{equation}\label{dis-kin8}
\frac{\partial h}{\partial t}+w\frac{\partial h}{\partial z}=u.
\end{equation}

The linear stability analysis of the above equations (\ref{dis-mom12}-\ref{dis-kin8}) is performed using $u=\tilde{u}e^{\Omega t+iqz}$, $w=\tilde{w}e^{\Omega t+iqz}$, $p=p_0+\tilde{p}e^{\Omega t+iqx}$, and $h=h_0+\tilde{h}e^{\Omega t+iqz}$. The linearisation of the momentum equations leads to
\begin{align}
  & \mu \left[ \frac{d}{dr}\left( \frac{1}{r}\frac{d(r\tilde{u})}{dr} \right)-{{q}^{2}}\tilde{u} \right]=\frac{d\tilde{p}}{dr}, \label{dis-lin9}\\ 
 & \mu \left[ \frac{1}{r}\frac{d}{dr}\left( r\frac{d\tilde{w}}{dr} \right)-{{q}^{2}}\tilde{w} \right]=iq\tilde{p}.\label{dis-lin10}
\end{align}
For the equation of mass conservation, we have
\begin{equation}\label{dis-lin11}
\frac{1}{r}\frac{d (r\tilde{u}) }{d r}+iq \tilde{w}=0.
\end{equation}

Using (\ref{dis-lin9}-\ref{dis-lin11}), the elimination of $\tilde{w}$ and $\tilde{p}$ leads to a fourth-order ordinary partial differential equation for $\tilde{u}$
\begin{equation}
\frac{d}{d r}\frac{1}{r}\frac{d}{dr}\left\lbrace r \frac{d}{dr}\left[\frac{1}{r}\frac{d (r\tilde{u})}{dr}\right]\right\rbrace-2q^2\frac{d}{dr}\left[\frac{1}{r}\frac{d(r\tilde{u})}{dr}\right]+q^4\tilde{u}=0.
\end{equation}
The general solution of this equation is \citep{cr2006}
\begin{equation} \label{dis-u}
\tilde{u}=C_1rK_0[qr]+C_2K_1[qr]+C_3rI_0[qr]+C_4I_1[qr],
\end{equation}
where $K_0 (K_1)$ and $I_0(I_1)$ are zeroth (first) order modified Bessel function of second and first kind. We can also get the expressions for $\tilde{w}$ and $\tilde{p}$ which are
\begin{align}
  & \tilde{w}=-\frac{1}{iq}\left\{ {{C}_{1}}\left[ 2{{K}_{0}}(qr)-qr{{K}_{1}}(qr) \right]-{{C}_{2}}q{{K}_{0}}(qr) \right. \nonumber \\ 
 & \left. \text{     }+{{C}_{3}}\left[ 2{{I}_{0}}(qr)+qr{{I}_{1}}(qr) \right]+{{C}_{4}}q{{I}_{0}(qr)} \right\},\label{dis-w} \\ 
 & \tilde{p}=2\mu \left[ {{C}_{1}}{{K}_{0}}(qr)+{{C}_{3}}{{I}_{0}}(qr) \right].\label{dis-p}
\end{align}
The four coefficients ($C_1$-$C_4$) are determined by the boundary conditions (\ref{dis-bou4}-\ref{dis-kin8}). For boundary conditions (\ref{dis-bou4}) and (\ref{dis-bou5}) at $r=a$, their linearised form are
\begin{equation}\label{dis-linb1}
\tilde{w}=\ell\frac{d \tilde{w}}{d r},
\end{equation}
\begin{equation}\label{dis-linb2}
\tilde{u}=0.
\end{equation}
And for boundary conditions (\ref{dis-nos6}-\ref{dis-kin8}) at $r=h_0$, their linearisation gives
\begin{equation}\label{dis-linb3}
\frac{d \tilde{w}}{d r}+iq\tilde{u}=0,
\end{equation}
\begin{equation}\label{dis-linb4}
-\tilde{p}+2\mu\frac{d \tilde{u}}{d {r}}=(-\gamma q^2+\gamma\frac{1}{h_0^2})\tilde{h},
\end{equation}
\begin{equation}\label{dis-linb5}
\Omega=\frac{\tilde{u}}{\tilde{h}}.
\end{equation}

A substitution of (\ref{dis-u}-\ref{dis-p}) into linearised boundary conditions (\ref{dis-linb1}-\ref{dis-linb5}) leads to  a set of four homogeneous equations, which is
\begin{equation}
\left( \begin{matrix}
   {{m}_{11}} & {{m}_{12}} & {{m}_{13}} & {{m}_{14}}  \\
   a{{K}_{0}}(qa) & {{K}_{1}}(qa) & a{{I}_{0}}(qa) & {{I}_{1}}(qa)  \\
   -{{K}_{1}}(q{{h}_{0}})+qh_0{{K}_{0}}(q{{h}_{0}}) & q{{K}_{1}}(q{{h}_{0}}) & qh_0{{I}_{0}}(q{{h}_{0}})+{{I}_{1}}(q{{h}_{0}}) & q{{I}_{1}}(q{{h}_{0}})  \\
   {{m}_{41}} & {{m}_{42}} & {{m}_{43}} & {{m}_{44}}  \\
\end{matrix} \right)\left( \begin{matrix}
   {{C}_{1}}  \\
   {{C}_{2}}  \\
   {{C}_{3}}  \\
   {{C}_{4}}  \\
\end{matrix} \right)=0,
\end{equation}
where the elements of first row are given by
\begin{align}
  & {{m}_{11}}=q(2\ell-a){{K}_{1}}(qa)-(\ell a{{q}^{2}}-2){{K}_{0}}(qa), \nonumber\\ 
 & {{m}_{12}}=-q{{K}_{0}}(qa)-\ell{{q}^{2}}{{K}_{1}}(qa), \nonumber\\
&{{m}_{13}}=-(\ell a{{q}^{2}}-2){{I}_{0}}(qa)-q(2\ell-a){{I}_{1}}(qa), \nonumber\\
&{{m}_{14}}=q{{I}_{0}}(qa)-\ell{{q}^{2}}{{I}_{1}}(qa).
\end{align}
The elements of fourth row are given by
\begin{align}
  & {{m}_{41}}=2\mu q{{h}_{0}}{{K}_{1}}(q{{h}_{0}})-D{{h}_{0}}{{K}_{0}}(q{{h}_{0}})/\Omega ,\nonumber \\ 
 & {{m}_{42}}=2\mu \left[ q{{K}_{0}}(q{{h}_{0}})+{{K}_{1}}(q{{h}_{0}})/{{h}_{0}} \right]-D{{K}_{1}}(q{{h}_{0}})/\Omega , \nonumber\\ 
 & {{m}_{43}}=-2\mu q{{h}_{0}}{{I}_{1}}(q{{h}_{0}})-D{{h}_{0}}{{I}_{0}}(q{{h}_{0}})/\Omega , \nonumber\\ 
 & {{m}_{44}}=-2\mu \left[ q{{I}_{0}}(q{{h}_{0}})-{{I}_{1}}(q{{h}_{0}})/{{h}_{0}} \right]-D{{I}_{1}}(q{{h}_{0}})/\Omega. 
\end{align}

Here $D$ is the driving force $D=\gamma ({{q}^{2}}-1/{{h}^{2}})$.

The vanishing of the determinant of $4\times4$ matrix  gives the dispersion relation $\Omega=\Omega(q)$. Numerically, we use Matlab to solve the determinant of the matrix. 
\section*{Appendix B: Capillary spectra from the Langevin model}
\setcounter{equation}{23}
For the Langevin equation formulated in the main text
\begin{equation}\label{eqn24}
\frac{\partial }{\partial{t}}\widehat{\delta h}=-\Omega \widehat{\delta h}+\frac{\zeta\Omega}{\beta} \widehat{N},
\end{equation}
its solution can be represented as the linear superposition of two contributions
\begin{equation}\label{eqn25}
\widehat{\delta h}={\widehat{\delta h}}_{\mathrm{det}}+{{\widehat{\delta h}}}_{\mathrm{flu}},
\end{equation}
where ${{\widehat{\delta h}}_{\mathrm{flu}}}$ is the contribution purely caused by thermal fluctuations and ${{\widehat{\delta h}}_{\mathrm{det}}}$ is the solution to the deterministic part of (\ref{eqn24}), i.e., $\frac{\partial \widehat{\delta h}}{\partial t}=-\Omega\widehat{\delta h}$; obtained as below 
\begin{equation}\label{eqn26}
\widehat{\delta {{h}}}_{\mathrm{det}}(q,t)=\widehat{\delta h}(q,0){{e}^{-\Omega t}},
\end{equation}
where the initial disturbance is $\widehat{\delta h}(q,0)$; here this is the Fourier transform of the liquid surface found in MD simulations at $t=0$.

To find the contribution of the fluctuating component to the spectrum, we determine the impulse response of the linear system $\frac{\partial \widehat{\delta h}}{\partial t}=-\Omega\widehat{\delta h}$ through:
\begin{equation} \label{eqn27}
\frac{\partial \widehat{\delta h}}{\partial t}=-\Omega\widehat{\delta h}+\delta. 
\end{equation}
Performing a Laplace transform of  (\ref{eqn27}) using $g(q,s)=\int_{0}^{\infty }{\widehat{\delta h}(q,t){e}^{-ts}d}t$ with zero initial disturbance $\widehat{\delta h}(q,0)=0$ gives
\begin{equation}
g=\frac{1}{s+\Omega},
\end{equation}
so that from the inverse Laplace transform, the impulse response is simply
\begin{equation}
H=\widehat{\delta h}={{e}^{-\Omega t}}.
\end{equation}
Now with thermal fluctuations $\zeta\widehat{N}$ as the input, we find
\begin{equation}\label{eqn30}
{{\widehat{\delta h}}_{\mathrm{flu}}}=\zeta\int_{0}^{t}{\widehat{N}\left( q,t-\tau  \right)}H(q,\tau )d\tau.
\end{equation}

As $\widehat{\delta h}$ is both a random and complex variable, the root mean square (r.m.s) of its norm is sought, which, from (\ref{eqn25}), is given by 
\begin{equation}
{{\left| \widehat{\delta h} \right|}_{\mathrm{rms}}}=\sqrt{\overline{{{\left| {{\widehat{\delta h}}_{\mathrm{det}}}+{{\widehat{\delta h}}_{\mathrm{flu}}} \right|}^{2}}}}  =\sqrt{{{\left| {{\widehat{\delta h}}_{\mathrm{det}}} \right|}^{2}}+\overline{{{\left| {{\widehat{\delta h}}_{\mathrm{flu}}} \right|}^{2}}}},
\end{equation}
(as the average of ${{\widehat{\delta h}}_{\mathrm{flu}}}$ is zero) where from (\ref{eqn26}) 
\begin{equation}
{{\left| {{\widehat{\delta h}}_{\mathrm{det}}} \right|}^{2}}={|\widehat{\delta h}(q,0)|}^2{{e}^{-2\Omega t}},
\end{equation}
and from (\ref{eqn30})
\begin{align}
  \overline{{{\left| {{\widehat{\delta h}}_{\mathrm{flu}}} \right|}^{2}}}&=\frac{\zeta^2\Omega^2}{\beta^2}\overline{{{\left| \int_{0}^{t}{\widehat{N}\left( q,t-\tau  \right)}H(q,\tau )d\tau  \right|}^{2}}} \nonumber \\ 
 & =\frac{\zeta^2\Omega^2}{\beta^2}\int_{0}^{t}{\overline{{{\left| \widehat{N}\left( q,t-\tau  \right) \right|}^{2}}}}H{{(q,\tau )}^{2}}d\tau \nonumber  \\ 
 & =\frac{\zeta^2\Omega^2}{\beta^2}\int_{0}^{t}{{{H}^{2}}}d\tau   \\ 
 & =\frac{\zeta^2\Omega}{2\beta^2}[1-{{e}^{-2\Omega t}}] \nonumber.
\end{align}
Thus, we obtain the spectrum of capillary waves as
\begin{align} \label{eqn34}
  S(q,t)&={{\left| \widehat{\delta h} \right|}_{\mathrm{rms}}} \nonumber  \\ 
 & =\sqrt{{|\widehat{\delta h}{{(q,0)}|}^{2}}{{e}^{-2\Omega t}}+S_s^2\left(1-{{e}^{-2\Omega t}} \right)} .
\end{align}
For an initially smooth surface, $\left|\widehat{\delta h}(q,0)\right|= 0$, (\ref{eqn34}) simplifies to 
\begin{align}
  S(q,t)=S_s\sqrt{\left( 1-{{e}^{-2\Omega t}} \right)},
\end{align}
which is \eqref{eqls} in the main text.
\section*{Appendix C: Measurements of slip length}
\begin{figure}
\includegraphics[width=\linewidth]{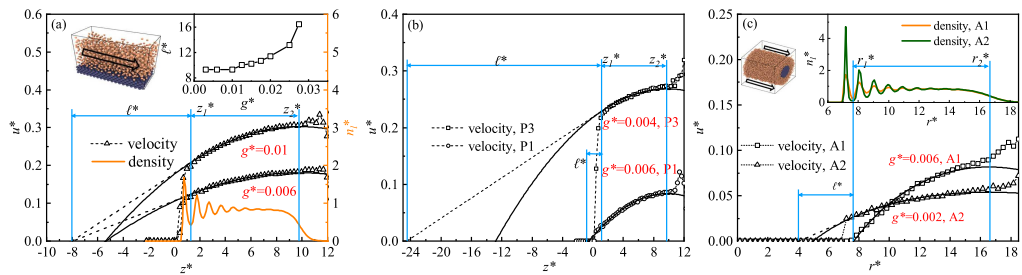}
\caption{Slip length measured using pressure-driven flows. Figures (a-b) are for planar films with (a) for case P2 and (b) for case P1 and P3. MD calculations of velocity (triangles) are fitted with analytical solutions (black solid lines) with the HB ($z_1$) at the first valley of MD density (yellow solid line) and FS ($z_2$) at $0.5n_l^*$. The inset shows slip length as a function of driving force. Figure (c) is for annular films, case A1 and A2. The inset shows the density profile.}  \label{figslip}
\end{figure}

Slip length is measured from independent configurations by simulating pressure-driven flow past a substrate surface as shown by the MD snapshots in the top-left corner of figure \ref{figslip}(a)  (for a planar film) and figure \ref{figslip}(c)  (for an annular film). The pressure gradient is created by applying a body force $g$ to the fluid. The generated velocity distribution is $u(z)=\frac{\rho g}{2\mu }(z-{{z}_{1}})(2{{z}_{2}}-z_1-z)+{{u}_{s}}$ for a planar film. Here $z_1$ and $z_2$ are positions of the hydrodynamic boundary (HB) and free surface (FS) for a planar film, respectively, and $u_s$ is the slip velocity at the HB. For an annular film, the axisymmetric velocity profile is $u(r)=-\frac{\rho g}{4\mu }\left[r^2-r^2_1-2r^2_1\log(r/r_2)\right]+{{u}_{s}}$, where $r_1$ and $r_2$ are positions of the HB and FS for this system.

The precise location of two boundary positions for each system is not trivial since there is an interfacial zone between the two different phases (solid-liquid and liquid-vapour) as demonstrated by the density distribution(the orange line in figure \ref{figslip}(a) and the inset of figure \ref{figslip}(c)). For the HB, research has showed it is located inside the liquid, between first-peak density and second-peak density rather than being located at the solid surface \citep{bo1994,ch2015} by comparing the analytical solution and MD measurements of the correlations of momentum density (an offset which matters when the interfacial layer has comparable width to the film). In line with this finding, we choose the position of HB at the first valley of density distribution: $z_{1}^*=1.3\sigma$ for a planar film and $r_{1}^*=7.65\sigma$ for an annular film (see figure \ref{figslip}(a) and figure \ref{figslip}(c)). The position of FS is determined in the standard way by the location of equimolar surface where density is $0.5n_l^*$, with $z_{2}^*=9.8\sigma$ for a planar film and $r_{2}^*=16.55\sigma$ for an annular film (see figure \ref{figslip}(a) and figure \ref{figslip}(c)). 

After locating the boundary, the slip velocity is obtained by fitting velocity profiles of MD data (symbols) with analytical expressions of velocity (solid black lines) as shown in figure \ref{figslip}(a). The slip length $\ell$ is the distance between the HB and the position where the the linear extrapolation of the velocity profile vanishes. Figure \ref{figslip}(a) is, in particular, for the measurement of slip length of case P2 where the slip length is measured to be $\ell^*=9.3\sigma$ (3.16 nm) ($\ell=0.68$ nm for P1 and $\ell=8.77$ nm for P3, see figure \ref{figslip}(b)). In figure \ref{figslip}(a), two different values of driving forces $g^*=0.01$ and $g^*=0.006$ are used to prove that the measured slip length is a constant independent of driving forces ($g^*\le 0.01$). However, as the inset shows, the slip length become force (shear)-dependent for $g^*\ge 0.01$, which is beyond current consideration \citep{th1997}. As the driving forces in the free-surface flows studied for capillary waves are small, the assumption of a constant slip length holds.
 
For annular films, as shown in figure \ref{figslip}(c), the slip lengths are $\ell=0$ nm (no-slip) for case A1 and $\ell=1.18$ nm for A2. Similar to the planar cases, we make sure that the slip lengths for annular cases are constant using driving forces with different strength.

We note that as the HB does not align with the edge of the solid, the effective thickness of the fluid domain simulated for capillary waves in the main text is different with its initial thickness. For a planar film, as the position of the initial free-surface is at $3.34$ nm (see \S\ref{sec2}) and the HB is at $z_1 = 0.44$ nm, the effective thickness of a planar film is 2.9 nm. For an annular film, this means $a = r_1 = 2.6$ nm and outer radius $h_0$ is 5.74 nm.

\bibliographystyle{jfm}

\end{document}